\documentclass[journal]{IEEEtran}
\IEEEoverridecommandlockouts

\usepackage[cmex10]{amsmath}

\usepackage{enumitem}
\usepackage{graphicx}
\usepackage{color}
\usepackage{amsmath}
\usepackage{mathtools}
\usepackage{multicol}
\usepackage{multirow}
\usepackage[english]{babel}
\usepackage{blindtext}
\usepackage{algorithm}
\usepackage{algorithmic}
\usepackage{balance}
\usepackage{amsfonts}
\usepackage{bm}
\usepackage{stfloats}
\usepackage{subfig}
\usepackage{amsthm}
\usepackage{amssymb}
\usepackage{setspace}
\usepackage[nosort]{cite}
\usepackage{CJK}
\usepackage{cite}
\usepackage{caption}
\usepackage{comment}
\usepackage{array,multirow}
\usepackage{graphicx}

\theoremstyle{plain}

\usepackage[table]{xcolor}

\begin{document}

\captionsetup[figure]{labelformat={default},labelsep=period,name={Fig.}}

\title{Spectrum Sharing in Satellite-Terrestrial Integrated Networks: Frameworks, Approaches, and Opportunities}

\author{Bodong Shang, Zheng Wang, Xiangyu Li, Chungang Yang, Chao Ren, and Haijun Zhang
\thanks{Bodong Shang is with Zhejiang Key Laboratory of Industrial Intelligence and Digital Twin, Eastern Institute of Technology, Ningbo, China, and also with the State Key Laboratory of Integrated Services Networks, Xidian University, Xi’an, China.}
\thanks{Zheng Wang, and Xiangyu Li are with Zhejiang Key Laboratory of Industrial Intelligence and Digital Twin, Eastern Institute of Technology, Ningbo, China.}
\thanks{Chungang Yang is with the State Key Laboratory of Integrated Services Networks, Xidian University, Xi’an, China.}
\thanks{Chao Ren and Haijun Zhang are with University of Science and Technology Beijing, China.}
}

% make the title area
\maketitle

\begin{abstract}
With the construction of low-earth orbit (LEO) satellite constellations, ubiquitous connectivity has been achieved. 
\textcolor{black}{Terrestrial networks (TNs), such as cellular networks, are mainly deployed in specific urban areas and use licensed spectrum. 
However, in remote areas where terrestrial infrastructure is sparse, licensed spectrum bands are often underutilized.
To accommodate the increasing communication needs, non-terrestrial networks (NTNs) can opportunistically access this idle spectrum to improve spectrum efficiency via spectrum sharing (SS).} 
Therefore, bringing NTNs to a shared spectrum with TNs can improve network capacity under reasonable interference management.
In satellite-terrestrial integrated networks (STINs), the comprehensive coverage of a satellite and the unbalanced communication resources of STINs make it challenging to manage mutual interference between NTN and TN effectively.
This article presents the fundamentals and prospects of SS in STINs by introducing four SS frameworks, their potential application scenarios, and technical challenges.
Furthermore, advanced SS approaches related to interference management in STINs and performance metrics of SS in STINs are introduced.
Moreover, a preliminary performance evaluation showcases the potential for sharing the spectrum between NTN and TN.
Finally, future research opportunities for SS in STINs are discussed.
\end{abstract}

\IEEEpeerreviewmaketitle

\section{Introduction}
Due to the growing need for broader and more reliable wireless network coverage and data transmission, non-terrestrial networks (NTNs) have emerged as a promising solution.
\textcolor{black}{In urban areas with dense users, the bandwidth shortage is a primary challenge, while in remote areas, the main problem is limited coverage. 
However, in such remote regions, a large portion of licensed terrestrial spectrum remains underutilized due to the sparse deployment of base stations \cite{sun2022integrated}. By allowing NTNs to opportunistically access these idle spectrum bands, spectrum sharing (SS) in STINs can significantly improve overall spectrum utilization without increasing allocation demands.}
Spectrum integration in satellite-terrestrial integrated networks (STINs) is a necessary step to improve the utilization efficiency of scarce spectrum resources.
However, the coexistence of NTNs and TNs can be challenging, particularly in terms of spectrum utilization and interference management. When NTN nodes, such as low-earth orbit (LEO) satellites, are incorporated into existing TNs via specific frequency bands, interference patterns can become more complex due to the different spatial propagation characteristics of signals. For example, the use of the Ka band or the Ku band for satellites may cause severe co-channel interference in existing communication systems. 
Moreover, within a fixed spectrum allocation, it is possible that while certain portions of the spectrum in NTNs are fully occupied, other segments within TNs may remain unutilized. The imbalance will lead to under-utilization of available spectrum resources and a significant hindrance to substantial improvements in system performance.

In recent years, the rapid growth of NTNs has paved the way toward realizing STINs. STINs, a promising candidate for future sixth-generation (6G) wireless communications, incorporate LEO nodes into existing TN to take advantage of NTN and meet diverse communication requirements~\cite{hui2025review}. 
By using STINs, the coverage, flexibility, and spectrum efficiency of the wireless network can be significantly improved, as terrestrial users can access various TN and NTN nodes from vast geographical areas. \textcolor{black}{Inevitably, the licensed spectrum resource will become scarcer in STIN as more NTN nodes are introduced. This encourages us to manage the licensed spectrum when TN and NTN co-exist in the STIN~\cite{li2023dynamic}}. Meanwhile, the licensed spectrum to the static TN nodes cannot always be utilized efficiently, as spatial spectrum holes cannot be eliminated within the geographic area of interest. Hence, mobile NTN users can admit access to users within the spectrum holes and promote spectrum efficiency~\cite{wei2023spectrum}. In existing work, one of the promising technologies for TN and NTN coexistence is SS, which allows the NTN to share the spectrum pre-allocated to the TN. SS allows the NTN to work in the same frequency band when the TN users are active, as long as its interference to the TN can be tolerated~\cite{wei2023spectrum,chen2023coverage}. In this way, the STIN can provide high-quality and seamless wireless services at a slight cost of performance loss under the practical restrictions of limited spectrum resources.

With the distinct advantages mentioned above, STINs can outperform conventional TNs in various practical fields. As the key technique in STINs, SS becomes the main bottleneck in network performance promotion and faces several challenges~\cite{heydarishahreza2024spectrum}. First, a universal SS architecture cannot always meet the diverse needs of the STIN. Therefore, the SS frameworks for all possible STIN transmission scenarios should be carefully designed, including the uplink and downlink of both TN and NTN. Second, since interference in STIN leads to system performance degradation, mitigating interference becomes the most critical issue. Generally, such mitigation can be implemented by isolating users from the interference source and carefully managing the TN and NTN spectrum~\cite{anjum2023space}. Meanwhile, the TN and NTN spectrum should be allocated appropriately, and the spectrum access scheme should be carefully designed~\cite{lee2023feasibility}. In addition, since STINs are hierarchical and heterogeneous, the proper selection of performance metrics for system evaluation is also crucial, depending on the design objectives.

\textcolor{black}{Apart from the above concerns, one factor that has a significant impact on SS is the framework in which the TN and NTN nodes operate, whether in uplink or downlink mode. In various frameworks, the mutual interaction between TN and NTN networks, as well as their transmission characteristics, may vary significantly, which are primary factors in designing proper SS strategies. Based on the SS frameworks, several state-of-the-art SS techniques are considered to address these concerns. For instance, a protection zone can be established to prevent NTN users within that zone from accessing the spectrum shared by TN and NTN. However, the defining protection zones remain a significant challenge. In addition, spectrum sensing is another crucial area for SS and is beneficial for STINs. Representative techniques, such as spatial spectrum sensing (SSS)~\cite{shang2020spectrum} and dynamic spectrum sensing~\cite{tang2020intelligent}, enable us to detect and reuse spatial and temporal spectrum holes within the STIN spectrum, respectively. In terms of spectrum allocation, game-theoretic models are competitive candidates, as they can represent TN and NTN as competing players~\cite{li2024game}. All of these topics warrant further study to address the performance requirements of future STINs.}

\captionsetup{font={scriptsize}}
\begin{figure*}[tp]
\begin{center}
\setlength{\abovecaptionskip}{+0.2cm}
\setlength{\belowcaptionskip}{+0.4cm}
\centering
  \includegraphics[width=7.1in, height=3.5in]{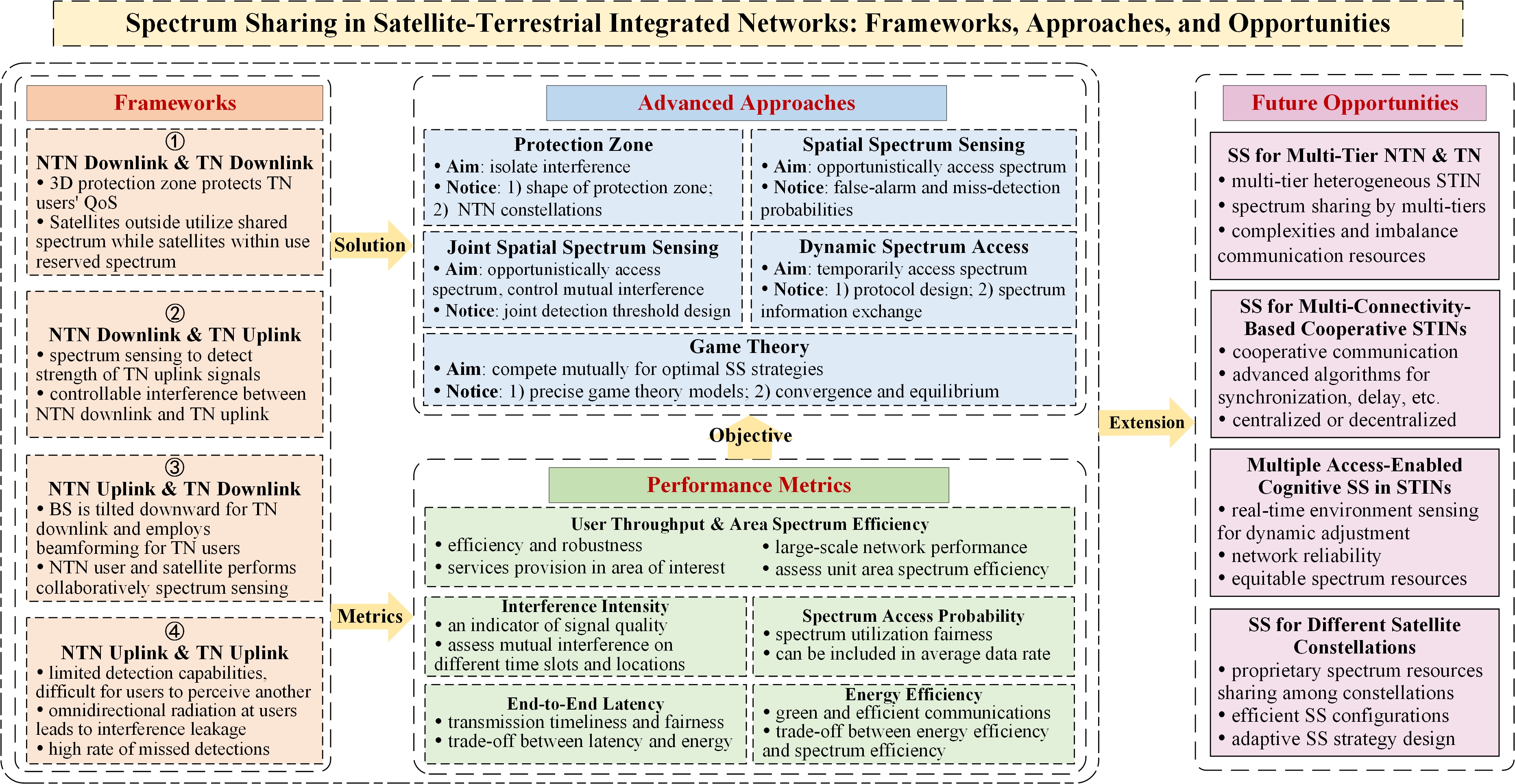}
\renewcommand\figurename{FIGURE}
\vspace{-4mm}
\caption{\scriptsize An overview of SS in STINs.}
\label{fig:table}
\end{center}
\vspace{-8mm}
\end{figure*}

This article focuses on the SS frameworks for various STIN scenarios and the state-of-the-art approaches to SS in STINs. The main contributions are summarized as follows.
\begin{itemize}
    \item We provide detailed discussions of possible SS frameworks in STIN and summarize their geographic and transmission characteristics, including all uplink and downlink scenario combinations for TN and NTN.
    \item We investigate several state-of-the-art approaches for possible SS in STINs to promote spectrum efficiency and system utilization from the spatial, temporal, and frequency domains. Detailed illustrations of these approaches are also provided.
    \item We summarize the potential performance metrics for evaluating the efficiency of SS in STINs and validate the improvement via a practical case study.  Finally, future research opportunities for SS in STINs are highlighted.
\end{itemize}

\section{Frameworks of SS in STINs}
\begin{figure*}[tp]
\begin{center}
\setlength{\abovecaptionskip}{+0.2cm}
\setlength{\belowcaptionskip}{+0.3cm}
\centering
  \includegraphics[width=6.8in, height=6.4in]{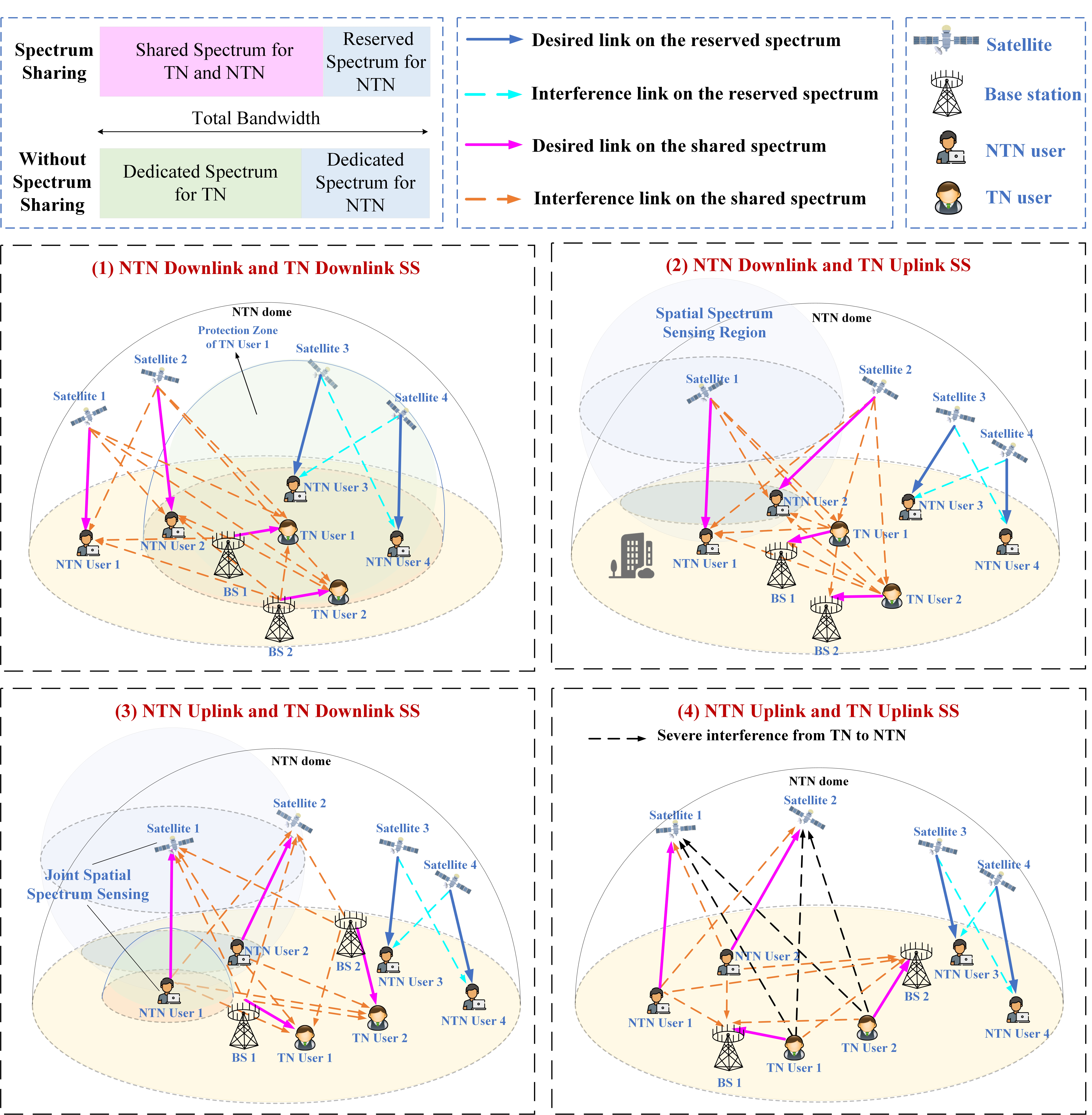}
\renewcommand\figurename{FIGURE}
\caption{\scriptsize Four frameworks of SS in STINs.}
\label{fig:frameworks}
\end{center}
\vspace{-8mm}
\end{figure*}

In this section, we present four frameworks of SS in STINs as illustrated in Fig. \ref{fig:frameworks}.
By implementing SS for NTN and TN in STINs, the total bandwidth is divided into two segments: 1) shared spectrum for both TN and NTN, and 2) reserved spectrum exclusively for NTN, as shown at the top of Fig. \ref{fig:frameworks}.

\captionsetup{font={scriptsize}}

\subsection{NTN Downlink and TN Downlink SS}
When a satellite transmits signals to an NTN user and a BS transmits signals to a TN user, they can access the shared spectrum, as shown in Fig. \ref{fig:frameworks}(1). For example, this framework corresponds to the NTN satellite n256 frequency band and new radio (NR) n65 and n66 frequency bands in the International Telecommunication Union Radiocommunication Sector (ITU-R) and the 3rd Generation Partnership Project (3GPP) \cite{3gpp.38.863}. It's worth noting that SS methods not only apply to the current spectrum regulations but also facilitate future spectrum utilization and standards development.
In this case, TN users can be regarded as primary users and NTN users as secondary users. Designing a three-dimensional (3D) protection zone for each TN user is essential. The protection zone can be either spherical or tapered. Satellites outside this protection zone may utilize the shared spectrum while those within continue to use the reserved spectrum.
This is because sharing spectrum between TN and NTN will not only enhance the communication performance of TN users through the expanded bandwidth but also result in interference.
The protection zone is primarily established to maintain the quality of service (QoS) for TN users. Theoretically, NTN users can also be affected by interference leakage from BSs, considering their beam directionality on the antenna surface in the horizontal plane. However, near-vertical transmissions from satellites can hardly be influenced. Moreover, satellites can mitigate mutual interference by constraining spectrum access within a specified distance range.

\subsection{NTN Downlink and TN Uplink SS}
Consider a scenario involving the sharing of a reverse link spectrum, in which an NTN user receives signals from a satellite. As depicted in Fig. \ref{fig:frameworks}(2), both the satellite and the NTN user can access the shared spectrum through TN uplink communications.
In this framework, satellite spectrum sensing can be performed to detect the strength of TN uplink signals, thereby reducing mutual interference.
If the uplink signal exhibits a low level of strength, the NTN downlink transmission can be operated to prevent significant mutual interference.
To be more specific, due to the directionality of satellite downlink signals, the corresponding interference to TN uplink transmissions on the near-horizontal plane is relatively small.
In addition, owing to the extended propagation distance and limited antenna gain associated with the satellite's side lobes, TN uplink users located within the reception range of these side lobes exert minimal interference on the satellite's spectrum sensing capabilities. 
In particular, spectrum sensing is conducted because satellites and NTN users cannot determine TN users transmitting in the current time slot or predict aggregated interference.

\subsection{NTN Uplink and TN Downlink SS}
When an NTN user transmits signals to a satellite, they can opportunistically access the shared spectrum with TN downlink communications, as shown in Fig. \ref{fig:frameworks}(3). For example, this framework corresponds to the NTN satellite n256 frequency band and NR n2, n25, and n70 frequency bands in ITU-R. 
This framework is feasible due to several factors. 
First, the BS is tilted downward for TN downlink communications and employs beamforming techniques to serve TN users. Due to the BS's antenna angle and beamforming, interference leakage from the BS to the satellite is minimized. 
Second, during NTN uplink communications, the NTN user and the satellite can collaboratively perform spectrum sensing. This allows them to opportunistically access the spectrum while minimizing interference to and from TN users.
It is important to note that, based on spectrum sensing conducted by NTN users, the aggregated interference power generated by NTN users is relatively lower than the desired signal power of the TN downlink communication. This is due to the high transmit power associated with TN BS, which also employs beamforming techniques.

\subsection{NTN Uplink and TN Uplink SS}
In the ITU-R, the NTN n256 frequency band and the NR n65 frequency band overlap. As shown in Fig. \ref{fig:frameworks}(4), allowing NTN uplink communications to share the spectrum with TN uplink communications poses several problems, necessitating that NTN use the n256 band in areas where TN does not use the n65 band.
First, numerous TN uplink users operate within a satellite's broad coverage area, and the distances between these users and NTN uplink users can be significant. This results in limited detection capabilities, making it difficult for either type of user to perceive the other effectively.
Second, TN users typically have a limited number of antennas, leading to considerable interference leakage from TN users in the vertical direction. This interference can overwhelm the desired signals from the NTN uplink communications.
Third, NTN users are likely to experience a high rate of missed detections regarding the presence of TN users. Consequently, this inability to accurately sense TN users can lead to significant interference from NTN uplink transmissions to TN communications. As a result, the communication performance of TN users often falls short of expectations. 
Sharing the spectrum between these two co-directional uplink links is unfavorable from the perspective of interference and sensing capabilities in many situations. However, this spectrum-sharing framework may be viable with appropriate spectrum-aware strategies when the number of TN uplink users is relatively small.

\section{Advanced Approaches to SS in STINs}
This section presents various approaches for achieving SS in STINs under different scenarios.

\captionsetup{font={scriptsize}}
\begin{figure*}[tp]
\begin{center}
\setlength{\abovecaptionskip}{+0.2cm}
\setlength{\belowcaptionskip}{+0.2cm}
\centering
  \includegraphics[width=7.0in, height=4.8in]{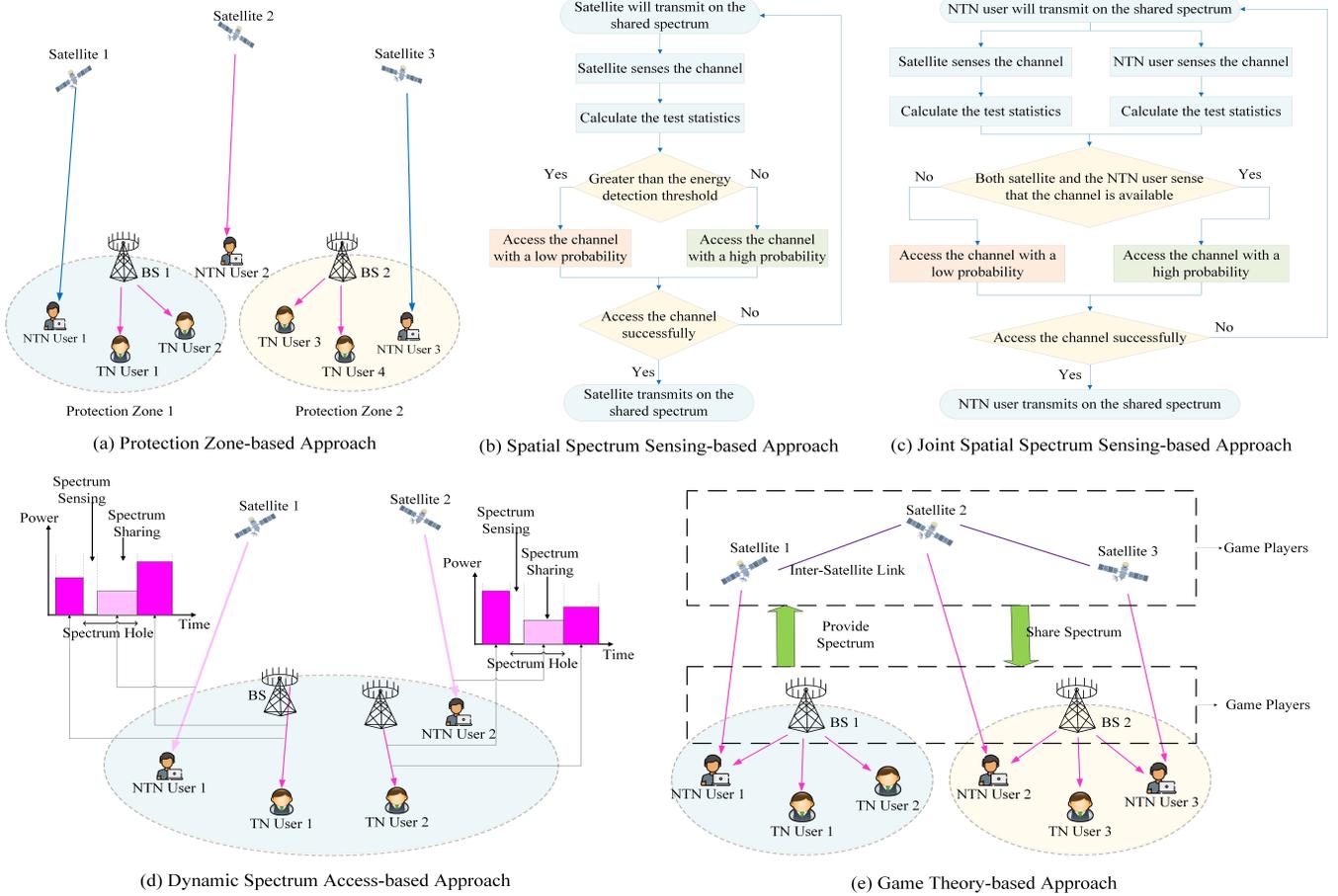}
\renewcommand\figurename{FIGURE}
\caption{\scriptsize Illustrations of advanced approaches to SS in STINs.}
\label{fig:approaches}
\end{center}
\vspace{-8mm}
\end{figure*}

\subsection{Protection Zone-Based Approach}

\textbf{Method:} 
Protection zones serve as an auxiliary approach in SS, as they define a geographical area in which users cannot access the spectrum of the primary system. SS techniques enable STIN users to sense and access the spectrum occupied by primary users to improve the spectrum efficiency. However, primary system interference cannot be ignored when users are close to interference sources. As protection zones are usually located around primary users, they can isolate interference for the users inside them, as shown in Fig. \ref{fig:approaches}(a). While the users are outside the protection zones, they are free to access the spectrum.

\textcolor{black}{\textbf{Applications:} With protection zones and user locations, the SS technique enables us to guarantee the key TN device performance and protect them from outage and malicious attacks. SS also enables intelligent monitoring of the spatial spectrum and spectrum sharing outside the protection zone.}

\textcolor{black}{\textbf{Difficulties and Solutions:} Since NTN is mobile and
provides flexible spectrum access, finding the protection zone boundary becomes the major challenge. The
dynamics of STIN architectures also impose another challenge on the protection zone. Both challenges require signal statistical modeling and analysis based on user distributions.}

\subsection{Spatial Spectrum Sensing (SSS)-Based Approach}
\textbf{Method:} SSS helps secondary users access the spectrum that relies on sensing rather than knowing the positions of interfering nodes \cite{9178984}. The term ``spatial'' refers to the ability of this method to capture the spatial characteristics of large-scale networks. The process begins with evaluating the interference power for a secondary user seeking to access a spectrum occupied by primary users. As shown in Fig. \ref{fig:approaches}(b), if the test statistic for the interference power exceeds a predefined energy detection threshold, it indicates that the channel is currently busy. In this case, the secondary user transmits with a low probability. Conversely, if the interference power is below the threshold, the secondary user transmits with a high probability.

\textcolor{black}{\textbf{Applications:} When SSS is implemented in STINs, NTN-related devices act as secondary users to monitor the busy status of the shared channel. By selecting an appropriate energy detection threshold, we can achieve efficient interference management from NTN toward TN. Meanwhile, we can provide spectrum information for terrestrial devices and thus support better service coverage.}

\textcolor{black}{\textbf{Difficulties and Solutions:} The major challenge of this method is precisely modeling signals and interference, which determines the false alarm and missed detection probabilities. It is crucial to control these probabilities following the Neyman-Pearson criterion. Possible solutions include calculating probability distributions based on interference statistical properties under various events or using machine learning techniques to fit these probabilities.}

\subsection{Joint Spatial Spectrum Sensing (JSSS)-Based Approach}
\textbf{Method:} JSSS involves two different types of network nodes working together to perform SSS. As shown in Fig. \ref{fig:approaches}(c), if both nodes determine that the channel is available for transmission, the secondary user will likely access the channel with a high probability. Conversely, if they disagree on availability, the secondary user will have a lower probability of accessing the channel.

\textcolor{black}{\textbf{Applications:} This approach is suitable for NTN uplink and TN downlink SS framework, as the satellite may experience significant interference from the TN when the number of ground base stations is high in specific areas. As a result, the SSS protects NTN communication from TN interference. Meanwhile, the SSS done by the NTN user helps manage the interference produced by the NTN towards the TN. In addition, SSS data analysis can help us find out abnormal historical data and better utilize the system spectrum resource.}

\textcolor{black}{\textbf{Difficulties and Solutions:} Since there are two different types of network nodes in JSSS, the joint decision whether SS can be accomplished has become the major challenge. To address this issue, the energy detection thresholds of the satellite and the NTN user can be jointly optimized under the Neyman-Pearson criterion.}

\subsection{Dynamic Spectrum Access-Based Approach}

\textbf{Method:} Dynamic spectrum access (DSA) is a widely adopted SS approach that takes advantage of the temporal holes of the licensed spectrum. In STINs, users can sense and access the available spectrum as long as they are not causing significant interference to other users or the spectrum is idle, as shown in Fig. \ref{fig:approaches}(d), and DSA can help achieve this objective by allowing secondary user access using the spectrum holes. In this way, the spectrum efficiency and system rate can thus be significantly promoted.

\textcolor{black}{\textbf{Applications:} Since NTNs have high mobility, STIN users can adopt DSA strategies to access the NTN spectrum. Similarly, NTNs can sense the spectrum licensed to TN users and adopt DSA to provide services for NTN users. Hence, flexible and robust networking in STINs can be achieved by adopting DSA. Meanwhile, DSA also enables flexible network access in remote areas where TN is no longer available.}

\textcolor{black}{\textbf{Difficulties and Solutions:} The sensing and decision mechanisms in DSA pose major challenges. To support these functionalities, the DSA protocol is also essential. Since NTN nodes need to detect spectrum holes at various geolocations over time, machine learning techniques incorporating dedicated geometric models are necessary for DSA in STINs. Meanwhile, since the NTN nodes are independent, the corresponding sensing schemes and protocols should also be designed.}

\subsection{Game Theory-Based Approach}

\textbf{Method:} Game theory is a powerful theoretical approach that models the interacting users as rational players aiming to maximize their benefits by selling and buying spectrum resources, as shown in Fig. \ref{fig:approaches}(e). In STINs, users tend to make their decisions and compete with each other to access the shared spectrum, ultimately deciding on the optimal SS pattern by reaching the game's Nash equilibrium.

\textcolor{black}{\textbf{Applications:} In STINs, the adoption of game theory models allows us to coordinate the interactions among TN and NTN nodes under potential spectrum conflict, thus maintaining TN and NTN performance as well as maximizing spectrum utilization. Game models can also be merged with DSA to decide the spectrum access patterns. This is suitable for NTN downlink and low-altitude communications, as the users can access the shared spectrum when deciding their access mode.}

\textcolor{black}{\textbf{Difficulties and Solutions:} Since STINs are hierarchical and heterogeneous, the internal coordination of various devices in SS often exhibits considerable complexity. To address the problem, selecting the appropriate network model and game model is a key issue in spectrum sensing, access, and management of STINs.}

\section{Performance Metrics of SS in STINs}
In this section, we introduce key metrics for evaluating the performance of SS in STINs from different perspectives.

\subsection{User Throughput and Area Spectrum Efficiency in STINs}
User throughput refers to the amount of data that the user can transmit and receive over a certain period, and system capacity is the sum of all user throughput. They both demonstrate the efficiency of STIN in continuously providing wireless service. Area spectrum efficiency (ASE) refers to the spectrum efficiency of a unit area, which can be used to evaluate the average performance of a large-scale network from a system-level perspective. Since STINs have dynamic architectures, a high and stable expected user throughput and ASE imply that STINs can provide high-quality wireless services, especially under large bandwidth and the assistance of SS.

\subsection{Interference Intensity between NTN and TN}
Interference intensity refers to the power strength per unit of bandwidth caused by secondary users to primary users, which denotes the negative impact of secondary users when accessing the shared spectrum in STINs. SS introduces additional interference into the system, as users experience interference from TN nodes during NTN downlink transmission, and TN nodes receive interference from users during NTN uplink transmission. Hence, a smaller interference intensity implies a higher efficiency and capacity of the STINs. Since the total interference power increases as the bandwidth grows, the STIN performance can be guaranteed by maintaining a satisfactory interference intensity.

 \subsection{Spectrum Access Probability of NTN User}
Spectrum access probability (SAP) refers to the likelihood that a secondary user will access the shared spectrum when they want to transmit packets. This metric indicates how fairly the spectrum is utilized among secondary users. When a channel is detected as busy, the SAP is low, prompting secondary users to either attempt to re-access the spectrum later or to utilize their reserved spectrum. For secondary users, such as NTN users or satellites within STINs, we can analyze or simulate the SAP when they employ an SS approach. Generally, a higher bandwidth is beneficial for SAP, as NTN users have more spectrum resources for SS.

\subsection{End-to-End Latency in STINs}
End-to-end (E2E) latency refers to the time it takes for a user to receive a data packet across the network, from the moment the packet is triggered. This metric is essential in the SS of STINs, as it reflects the timeliness and fairness of packet transmission. In STINs, secondary users may need to delay their transmissions if they cannot access the spectrum in the current time slot. Additionally, when dealing with satellite-to-ground communication links, it is essential to consider propagation delay, which can add several milliseconds to the E2E latency. Generally, a larger bandwidth indicates a higher data rate, which ultimately results in a smaller E2E latency.

\subsection{Energy Efficiency of NTN and/or TN}
Energy efficiency (EE) refers to the number of bits a user can transmit by consuming per unit of power. As a key performance metric, EE is calculated by dividing the data throughput by the power consumption, which represents the energy utilization in transmitting signals. Since NTN nodes and users in STINs often face limited energy supply restrictions, it is crucial to promote their EE to transmit more data and provide better QoS. Meanwhile, increasing EE also enables us to better utilize scarce spectrum resources. Thus, a good guarantee of EE is essential for the overall system performance of STINs, and the SS technique can naturally promote EE, especially under large bandwidth conditions.

\section{Case Study: SS in STINs Using Protection Zone-based Approach}

\captionsetup{font={scriptsize}}
\begin{figure}[tp]
\begin{center}
\setlength{\abovecaptionskip}{+0.2cm}
\setlength{\belowcaptionskip}{-0.4cm}
\centering
  \includegraphics[width=3.0in, height=5.2in]{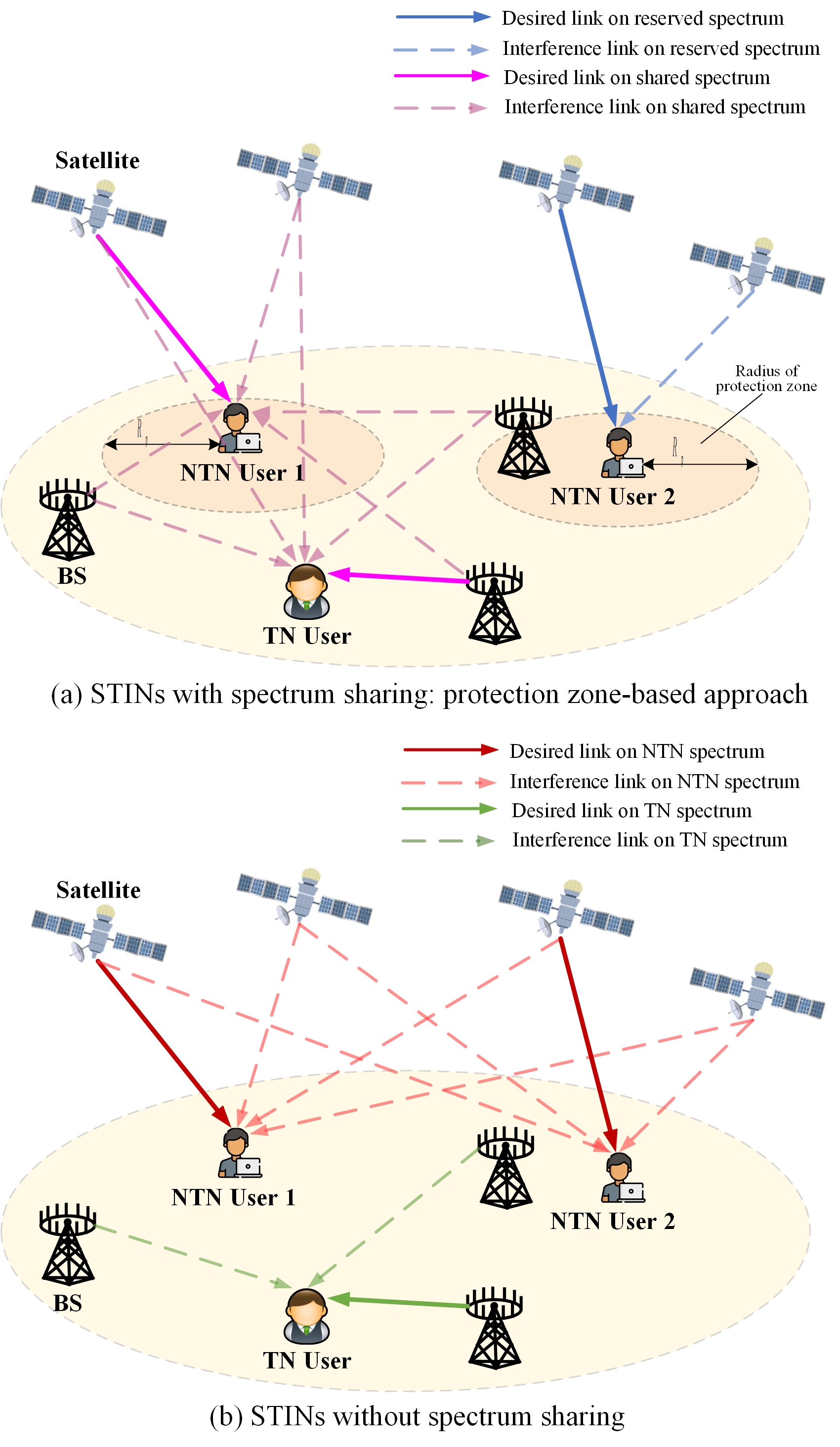}
\renewcommand\figurename{FIGURE}
\caption{\scriptsize Simulation scenario.}
\label{fig:Simulation_scenario}
\end{center}
\vspace{-4mm}
\end{figure}

\captionsetup{font={scriptsize}}
\begin{figure}[tp]
\begin{center}
\setlength{\abovecaptionskip}{+0.2cm}
\setlength{\belowcaptionskip}{-0.4cm}
\centering
  \includegraphics[width=3.0in, height=5.2in]{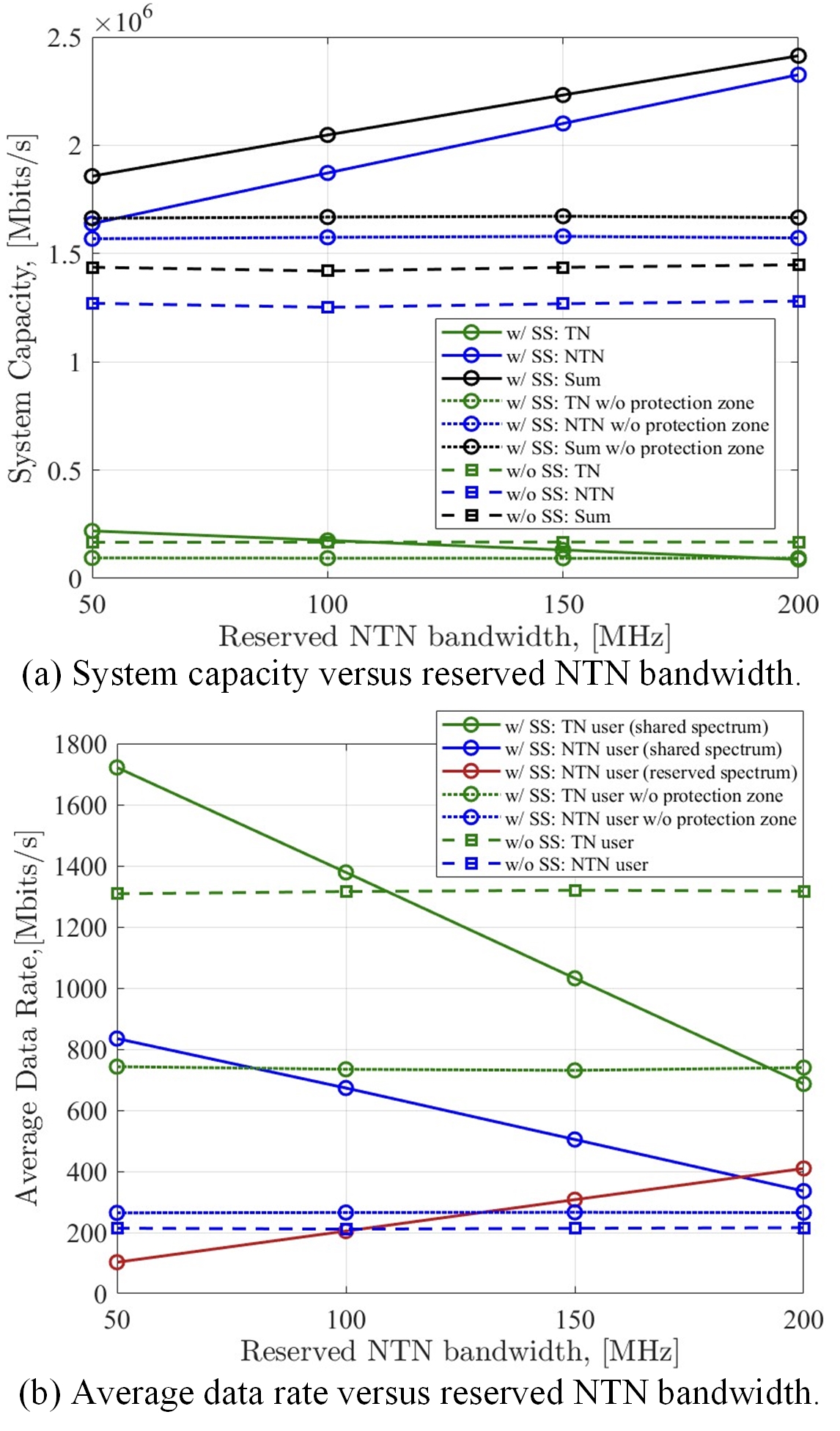}
\renewcommand\figurename{FIGURE}
\caption{\scriptsize System capacity and average data rate of STINs with SS and without SS. The total bandwidth is $300$ MHz, while the NTN bandwidth and the TN bandwidth are fixed to $120$ MHz and $180$ MHz, respectively, for STINs without SS. The BSs are randomly distributed in a $200$ km$\times 200$ km remote area with a density of $\lambda_\text{T} = 3\times 10^{-3}$/km$^2$. Each BS has a service radius of $R_\text{T} = 10$ km and transmit power of $P_\text{T} = 46$ dBm. The pathloss exponent in TNs is $\alpha_\text{T} = 3.5$. 
A typical TN user is assumed to be located at the center of the serving area, while typical NTN users are $50$ km away from the typical TN user.
The noise power at the user side is $-110$ dBm. 
In NTNs, the satellites are distributed on an orbital sphere with a density of $\lambda_\text{N} = 1\times 10^{-5}$/km$^2$ at an altitude of $500$ km. The transmit power, main-lobe gain, side-lobe gain, and path loss exponent are $P_\text{N} = 50$ dBm, $G_\text{ml} = 30$ dBi, $G_\text{sl} = 10$ dBi, and $\alpha_\text{N} \approx 2.0$, respectively.}
\label{fig:Simulation_results}
\end{center}
% \vspace{-3mm}
\end{figure}

In this section, we quantitatively study the effectiveness of SS in STINs by adopting the protection zone-based approach. The simulation scenarios are first depicted in Fig. \ref{fig:Simulation_scenario}. 
The NTN downlink and TN downlink framework is considered, where the NTN users are the primary users and the TN users are the secondary users.
A protection zone-based approach is adopted, so that the BSs are located at least $R_\text{p}$ away from NTN users.
For the case of SS in Fig. \ref{fig:Simulation_scenario}(a), NTN users are primary users, each having a protection zone with radius $R_\text{p}$. An NTN user will use the reserved spectrum if there is a BS inside the protection zone and will use the shared spectrum if there is no BS inside. If SS is not adopted in STINs as shown in Fig. \ref{fig:Simulation_scenario}(b), NTN users and TN users will experience interference from the NTN spectrum and TN spectrum, respectively.
This case study evaluates the aforementioned networks using two key performance metrics. First, the average data rate is determined by multiplying the product of spectral efficiency and bandwidth. 
Second, the system capacity of the NTN (or TN) is assessed by multiplying the spectral efficiency by the satellite (or BS) density, the surface area of satellites (or BSs), and the available bandwidth.

The simulation results are shown in Fig. \ref{fig:Simulation_results}, comparing the following three scenarios, i.e., \textbf{Scenario 1}: SS with the protection zone-based approach; \textbf{Scenario 2}: SS without the protection zone-based approach; \textbf{Scenario 3}: the setting without SS.
Note that, as reserved NTN bandwidth is considered in Scenarios 2 and 3, the system capacity and average data rate of these scenarios remain almost the same. 
Specifically, in Fig. \ref{fig:Simulation_results}(a), the TN system capacity in Scenario 1 first surpasses that of Scenario 2 when $B_\text{R}$ is around $100$ MHz, but the latter subsequently exceeds it. 
After applying SS in Scenario 2, the excessive bandwidth can be utilized more efficiently, leading to an increase in the sum and NTN system capacity compared to Scenario 3.
The protection zone-based approach in Scenario 1 can utilize excessive reserved bandwidth and decrease co-frequency interference by setting an empirical protection radius $R_\text{p}$. If a BS is located within distance $R_\text{p}$, the NTN user will switch to the shared spectrum instead of the reserved spectrum. A gradual increase in both the sum and the NTN system capacity can be observed.

In Fig. \ref{fig:Simulation_results}(b), the average data rate of the users in STINs with SS is compared to that of the users in STINs without SS. 
Compared to Scenario 3, the NTN user will have a higher data rate, but the TN user will have a lower data rate in Scenario 2, primarily due to more substantial co-frequency interference from satellite networks.
In Scenario 1, with an increase of $B_\text{R}$, the data rate of the NTN user and the TN user using shared spectrum decreases because more bandwidth is allocated to the reserved spectrum. However, the data rate of the NTN user using the reserved spectrum increases. 
Therefore, by selecting the appropriate reserved bandwidth according to implementation requirements, a win-win solution can be obtained by using SS with the protection zone design.

\section{Future Opportunities of SS in STINs}
\begin{figure*}[tp]
\begin{center}
\setlength{\abovecaptionskip}{+0.2cm}
\setlength{\belowcaptionskip}{+0.2cm}
\centering
  \includegraphics[width=6.6in, height=4.2in]{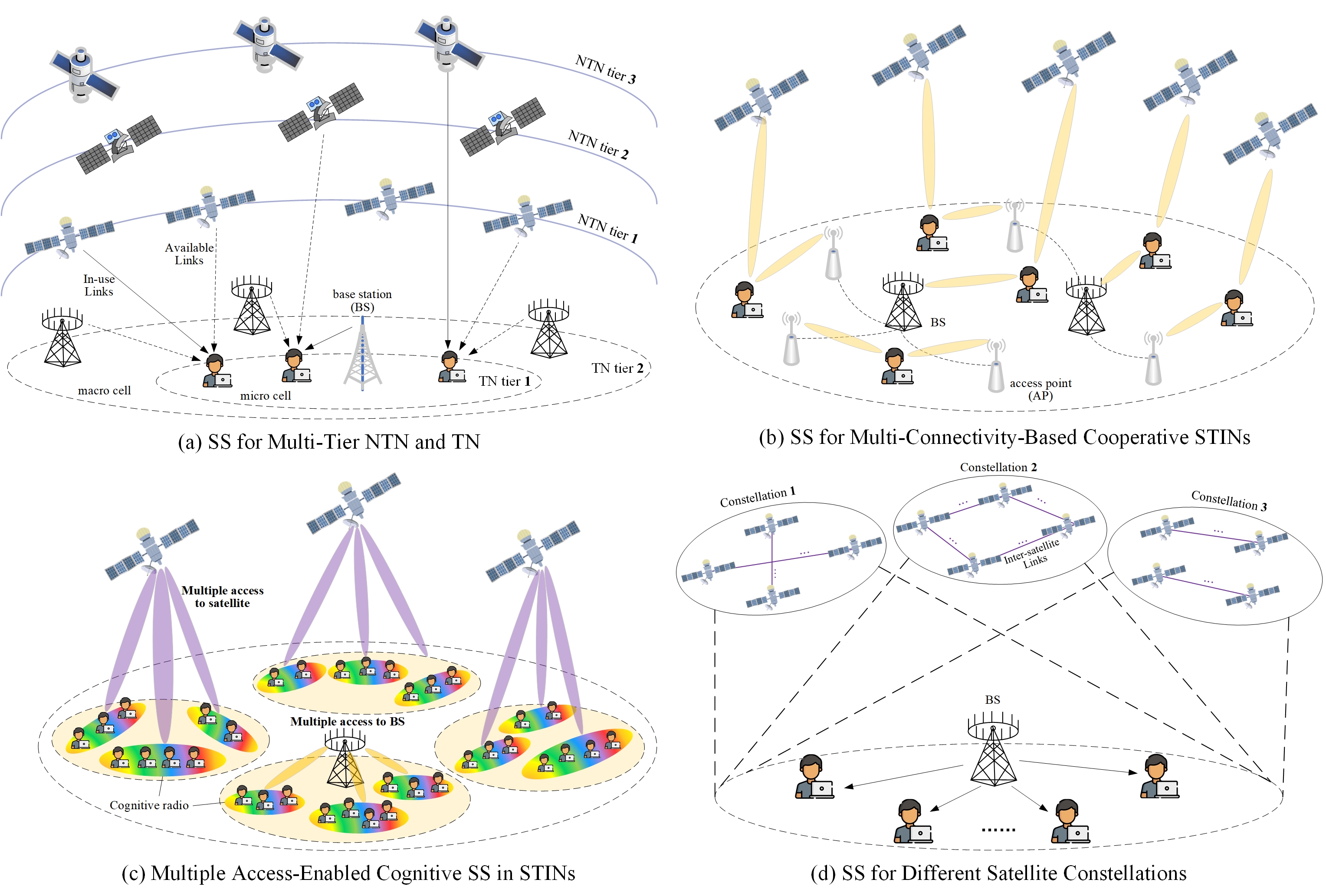}
\renewcommand\figurename{FIGURE}
\caption{\scriptsize Future opportunities of SS in STINs.}
\label{fig:opportunities}
\end{center}
\vspace{-8mm}
\end{figure*}

In this section, we briefly discuss future applications of SS in STINs. An overview of these potential applications is presented in Fig. \ref{fig:opportunities}, and detailed descriptions of each are provided below.

\captionsetup{font={scriptsize}}

\subsection{SS for Multi-Tier NTN and TN}
Due to the limited availability of orbital positions, satellites are deployed at various altitudes. Different types of satellites have varying antenna patterns and transmit powers, and the number of satellites in different tiers also varies. As a result, the single-tier STIN will evolve into a multi-tier heterogeneous STIN (HetSTIN). 
However, this multi-tier HetSTIN faces a significant challenge: the conflict between limited spectrum resources and the increasing demand for satellite services. Effectively sharing the spectrum among the multi-tier NTN and multi-tier TN is crucial for successfully integrating satellite and terrestrial networks.
Addressing the complexities of interference management and the imbalanced communication resources that arise from the heterogeneous nature of multi-tier HetSTINs is essential in the design of SS.

\subsection{SS for Multi-Connectivity-Based Cooperative STINs}

Multi-connectivity in STINs, as referred to in the 3GPP Technical Standard \cite{3gpp.38.821}, enables a user to simultaneously connect to multiple network nodes, including BSs or access points in TNs and satellites in NTNs. 
With proper synchronization and mitigation of delay, phase misalignment, and Doppler shifts using algorithms \cite{kojima2023timing}, cooperative communication, such as relaying, joint transmission, or reception, can be utilized to connect different network tiers in STINs and promote more resilient and adaptable communications.

However, the bandwidth allocated to a user, which may have multiple connection links, can be restricted due to the scarcity of available spectrum resources. 
Hence, the allocation of spectrum resources among multiple users, each possessing multiple links, presents significant research opportunities. Additionally, the organization of cooperative network nodes, whether centralized or decentralized, warrants further exploration and investigation.

\subsection{Multiple Access-Enabled Cognitive SS in STINs} 
\textcolor{black}{
Multiple access allows multiple users to share the same spectrum simultaneously. The cognitive radio can sense the real-time radio environment and adjust the transmission parameters accordingly. With these sensing results, secondary users can assess the unutilized spectrum without interfering with primary users. 
For instance, non-orthogonal multiple access (NOMA) allows users to share the same frequency resources with different power levels. This can be paired with spectrum sensing to adjust user clustering and power allocation based on the levels of interference. Rate-splitting multiple access (RSMA) can flexibly split messages into common and private parts, which can be adapted to the variable interference landscape in shared spectrum environments.
}

\textcolor{black}{
Using these multiple access schemes, STINs can not only increase spectrum efficiency but also support user fairness and low-latency requirements under dynamic spectrum availability. Future research can focus on jointly optimizing spectrum sensing, access control, and user scheduling for each multiple-access scheme.
}

\subsection{SS for Different Satellite Constellations}
As satellite constellations have expanded to mega size in recent years, the number of different satellite constellations built by different satellite companies is also increasing. Satellite constellations can also share their proprietary spectrum resources to improve the system's capacity. Due to the mobile nature of the satellites, the NTN architecture is always dynamic. It changes rapidly, posing a significant challenge to maintaining reliable and high-quality services under the constraints of limited and congested spectrum resources. Different satellite constellations face different transmission conditions as service providers. They require compatible spectrum access strategies. SS techniques enable the high-efficiency utilization of spectrum resources while introducing controllable interference. Meanwhile, to meet the constellation dynamics, the designed SS strategy should also be adaptive to the availability of satellites and spectrum resources over time in providing wireless services.

% \vspace{-2mm}
\section{Conclusions}
This article investigated the frameworks, approaches, performance metrics, and future research opportunities for SS in STINs. We introduced four general SS frameworks in STINs, including different uplink and downlink combinations for NTN and TN, and discussed their characteristics. Then, state-of-the-art approaches for SS to reuse spectrum resources in space, time, and frequency domains in STINs were comprehensively reviewed. More importantly, related metrics were presented to study the SS techniques further and evaluate the system's and individual users' performance. A case study on the protection zone-based approach for SS in STINs was conducted, and the simulation results demonstrated that a win-win solution could be achieved for enhanced system capacity. Finally, we provided promising opportunities and insightful future research on SS in STINs.

% \vspace{-2mm}
\bibliographystyle{IEEEtran}
\bibliography{references.bib}

\section*{Biographies}

\vspace{-14mm}

\begin{IEEEbiographynophoto}{Bodong Shang}
(bdshang@eitech.edu.cn) 
received his Ph.D. degree from the Department of Electrical and Computer Engineering at Virginia Tech, Blacksburg, USA. He was a Postdoctoral Research Associate at Carnegie Mellon University, Pittsburgh, USA. He held a research internship position at Nokia Bell Labs, USA. He is an Assistant Professor at Eastern Institute for Advanced Study, Eastern Institute of Technology (EIT), Ningbo, China. His research interests include space-air-ground-sea integrated networks, non-terrestrial networks, and space information networks.
\end{IEEEbiographynophoto}

\vspace{-14mm}

\begin{IEEEbiographynophoto}{Zheng Wang}
(zwang@idt.eitech.edu.cn) 
received his Ph.D. degree from the Department of Electrical and Computer Engineering at George Mason University, Fairfax, USA, in 2020, and he was a Postdoctoral Research Associate at Virginia Tech, Blacksburg, USA. He is a Research Associate Professor at Ningbo Institute of Digital Twin, Eastern Institute of Technology (EIT), Ningbo, China. His research areas include satellite-terrestrial integrated networks, game theory, and physical layer communication.
\end{IEEEbiographynophoto}

\vspace{-14mm}

\begin{IEEEbiographynophoto}{Xiangyu Li}
(xyli@eitech.edu.cn) 
received an M.S. degree in Electrical and Computer Engineering from Georgia Institute of Technology, Atlanta, USA, in 2023, and he is currently pursuing a Ph.D. degree with Shanghai Jiao Tong University (SJTU), Shanghai, China, at the Eastern Institute of Technology (EIT), Ningbo and SJTU Joint PhD Program. His research interests include space-air-ground integrated networks, non-terrestrial networks, and performance analysis of wireless systems.
\end{IEEEbiographynophoto}

\vspace{-14mm}

\begin{IEEEbiographynophoto}{Chungang Yang}
(guideyang2050@163.com) 
is a Professor at Xidian University, leading the GUIDE: Game, Utility, Artificial Intelligent Design for Emerging Communications research team. His research interests include artificial intelligence, 6G wireless mobile networks, intent-driven networks, space-terrestrial networks, and game theory for emerging communication networks.
\end{IEEEbiographynophoto}

\vspace{-14mm}

\begin{IEEEbiographynophoto}{Chao Ren}
(chaoren@ustb.edu.cn) 
is an Associate Professor and Deputy Director with the University of Science and Technology in Beijing, China. He is an Editorial Board Member for the Journal of Communications and Frontiers in Computer Science, recognized with an Outstanding Editor Award. He has been a Guest Editor for Space-Integrated-Ground Information Networks and Information Security Research, Co-Chair of the UCOM 2023-2025 Workshop, and Technical Program Committee member for over 30 international conferences in communications.
\end{IEEEbiographynophoto}

\vspace{-14mm}

\begin{IEEEbiographynophoto}{Haijun Zhang}
(haijunzhang@ieee.org) 
is a Professor and dean at the University of Science and Technology in Beijing, China. He serves/served as an Editor of IEEE Transactions on Information Forensics and Security, IEEE Transactions on Communications, IEEE Transactions on Network Science and Engineering, and IEEE Transactions on Vehicular Technology. He received the IEEE CSIM Technical Committee Best Journal Paper Award in 2018, the IEEE ComSoc Young Author Best Paper Award in 2017, and the IEEE ComSoc Asia-Pacific Best Young Researcher Award in 2019. He is a Fellow of IEEE.
\end{IEEEbiographynophoto}

\end{document}